\documentclass[a4paper,fleqn]{cas-dc}

\usepackage[authoryear,longnamesfirst]{natbib}
\usepackage{subfig}
\setcitestyle{numbers,open=[,close=]}

\def\tsc#1{\csdef{#1}{\textsc{\lowercase{#1}}\xspace}}
\tsc{WGM}
\tsc{QE}

\newcommand{\beginsupplement}{%
        \setcounter{table}{0}
        \renewcommand{\thetable}{S\arabic{table}}%
        \setcounter{figure}{0}
        \renewcommand{\thefigure}{S\arabic{figure}}%
     }

\begin{document}
\let\WriteBookmarks\relax
\def\floatpagepagefraction{1}
\def\textpagefraction{.001}

\shorttitle{Multimodal N-of-1 trials: A Novel Personalized Healthcare Design}    

\shortauthors{Jingjing Fu, Shuheng Liu, Siqi Du, Siqiao Ruan, Xuliang Guo, Weiwei Pan, Abhishek Sharma, Stefan Konigorski}  

\title [mode = title]{Multimodal N-of-1 trials: A Novel Personalized Healthcare Design}  

\tnotemark[Multimodal N-of-1 trials: A Novel Personalized Healthcare Design] 


%

\author[1]{Jingjing Fu}
\author[1]{Shuheng Liu}
\author[1]{Siqi Du}
\author[1]{Siqiao Ruan}
\author[1]{Xuliang Guo}
\author[1]{Weiwei Pan}
\author[1]{Abhishek Sharma}
\author[2,3,4]{Stefan Konigorski}[
    orcid=0000-0002-9966-6819
]
\ead{stefan.konigorski@hpi.de}


\cormark[1]









\affiliation[1]{organization={School Of Engineering And Applied Sciences, Harvard University},
            state={Cambridge, Massachusetts},
            country={USA}}
\affiliation[2]{organization={Department of Statistics, Harvard University},
            state={Cambridge, Massachusetts},
            country={USA}}
\affiliation[3]{organization={Digital Health Center, Hasso Plattner Institute for Digital Engineering, University of Potsdam},
            state={Potsdam},
            country={Germany}}
\affiliation[4]{organization={Hasso Plattner Institute for Digital Health at Mount Sinai, Icahn School of Medicine at Mount Sinai},
            state={New York, NY},
            country={USA}}


\cortext[1]{Corresponding author: Stefan Konigorski, Department of Statistics, Harvard University, 150 Western Avenue, Boston 02134, Massachusetts, USA.}



\begin{abstract}
N-of-1 trials aim to estimate treatment effects on the individual level and can be applied to personalize a wide range of physical and digital interventions in mHealth. In this study, we propose and apply a framework for multimodal N-of-1 trials in order to allow the inclusion of health outcomes assessed through images, audio or videos. We illustrate the framework in a series of N-of-1 trials that investigate the effect of acne creams on acne severity assessed through pictures. For the analysis, we compare an expert-based manual labelling approach with different deep learning-based pipelines where in a first step, we train and fine-tune convolutional neural networks (CNN) on the images. Then, we use a linear mixed model on the scores obtained in the first step in order to test the effectiveness of the treatment. The results show that the CNN-based test on the images provides a similar conclusion as tests based on manual expert ratings of the images, and identifies a treatment effect in one individual. This illustrates that multimodal N-of-1 trials can provide a powerful way to identify individual treatment effects and can enable large-scale studies of a large variety of health outcomes that can be actively and passively assessed using technological advances in order to personalized health interventions.
\end{abstract}




\maketitle
\section{Introduction}
Clinical trials, in particular randomized controlled trials (RCTs), are the gold standard study design for evaluating the effectiveness of medical, behavioral or mHealth interventions on a population level. As such, clinical trials play a vital role in the development of health interventions and have a significant impact both on medical research and on healthcare.
In RCTs, participants are randomly assigned to either the treatment group, which receives the treatment to be tested, or the control group, which receives a placebo or standard treatment. With the random assignment of participants, RCTs can control for confounding variables that may affect the study. 
Through their population-level approach, the results of RCTs reflect the effectiveness of treatment on an average level. The most effective treatment on average in the population, however, might not work at all for a given individual, so for personalizing treatments, other approaches are needed.

N-of-1 trials are individual multi-crossover trials designed to study the effectiveness of treatments on a personal level \cite{Nikles2015book}. In contrast to RCTs, each participant in an N-of-1 trial is viewed as an independent system and receives treatments sequentially over multiple periods. Thereby, N-of-1 trials allow inference on personalized treatment effects. There have been many applications of N-of-1 trials in the medical field. For example, recent research has shown that  N-of-1 trials can be applied to evaluate the side effects of statins on muscle symptoms \citep{herrett2021statin}, which allows for more personalized guidance of treatment. Another example is in pediatric oncology 
\citep{kyr2021n} where an N-of-1 trial has been applied to address the problem of limited patient samples, facilitating the development and approval of new drugs.

Performing N-of-1 trials digitally allows to evaluate mHealth applications on an individual level and to personalize them. At the same time, digital N-of-1 trials offer the potential to make trials available to patients and users worldwide and scale their applicability. To this aim, we have developed an online open-source platform named StudyU in previous work \cite{konigorski2022studyu}. With this platform, physicians can utilize a collaborative web application to design and perform digital N-of-1 trials, and patients can participate in these trials to find out whether a particular treatment is effective on themselves, by providing self-reported health outcomes. To fully use the capabilities of smartphone-based evaluation of (m)Health interventions in N-of-1 trials, we propose to collect multimodal health outcomes, such as images, audio or videos recorded by the study participants on their phones. Designing, running, and analyzing the data of such trials requires novel design considerations and novel methodology that is not yet available.


In this paper, we propose a first-of-its-kind framework to conduct multimodal N-of-1 trials that leverages user-recorded images as health outcomes, which is an important milestone to perform fully-flexible powerful digital N-of-1 trials. For the process of data collection, we propose a protocol that specifies lighting, device and direction to guarantee data quality. We propose a 2-step pipeline for the modeling procedure evaluating the effectiveness of treatments. In a first step,  we train and fine-tune a convolutional neural network (CNN) on the images. Then, we use a linear mixed model on the scores obtained in the first step in order to test the effectiveness of the treatment. We illustrate this framework in a toy series of N-of-1 trials on acne severity in order to demonstrate the validity of our proposed pipeline. The results show that the CNN-based test on the images provides a similar conclusion as tests based on manual expert ratings of the images, and identifies a treatment effect in one individual. Importantly, our proposed framework can be automated and opens up many possibilities to incorporate more data modalities into digital N-of-1 trials.








\section{Motivating Illustration: Multimodal N-of-1 Trials of Acne}

\subsection{Overview}

In order to illustrate our proposed framework for multimodal N-of-1 trials, we use a motivating example study where the aim was to evaluate the effect of acne creams on acne, which is a common skin condition. 
There are many products in the market that have been developed to improve acne conditions. The study examined two products -- a \textit{benzoyl peroxide based acne cream}, hereinafter known as product A, and a \textit{salicylic acid based acne control gel}, hereinafter known as product B. 
The de-identified data of this toy series of N-of-1 trials has been previously published and is publicly available at \url{https://github.com/HIAlab/Multimodal_Nof1/tree/main/Data/Acne_Nof1_trial}. The application of our framework in the following sections is secondary analysis of this de-identified data, i.e. it does not constitute human subjects research and is hence exempt from ethics approval.

\subsection{Treatment Schedule}
\begin{figure}[h]
    \includegraphics[width=0.45\textwidth]{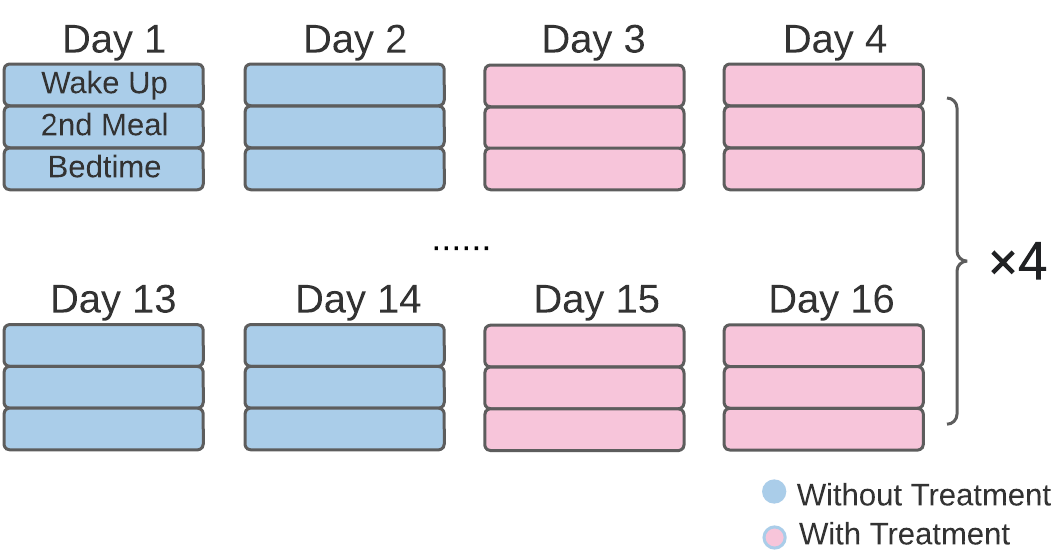}
    \caption{Experiment setup with 4 blocks, each including 2 days with treatment A and 2 days with treatment B in alternating design.}
    \label{fig:experiment-setup}
\end{figure}

The series of N-of-1 trials was conducted on a total of five individual participants -- two participants using product A and three participants using product B -- over a period of 16 days. Each participant applied the treatment according to a specified schedule on spots of facial or body acne. For each participant, images were taken three times a day for a continuous period of 16 days – right after waking up, right after second meal of day, and right before bedtime, respectively. The time to collect data was dependent on the daily schedule of each participant to allow for comparable results across participants. The 16-day study consisted of 4 four-day blocks, where each participant collected data first for two days without treatment and then two days with treatment. See Figure \ref{fig:experiment-setup} for an illustration and see Table \ref{tab:schedule} for a complete schedule. This yielded 48 collected images per participant. There was no missing data.

\subsection{Image Protocol Specification}
In order to ensure data quality, the study participants followed a strict set of protocols for taking the images.

\begin{enumerate}
    \item Room Settings: Every photo was taken in the same position in an enclosed room with fixed lighting. There was no natural light in the room.
    \item Device: For each participant, the device used for taking photos was the same handheld smartphone with the same optical settings.
    \item  Angle and Distance: For each participant, the photos were taken with the lens directly facing at the area of interest and at a distance of around 10 centimeters to guarantee uniformity of collected data.
\end{enumerate}

See Figures \ref{fig:xg-sample} - \ref{fig:sd-sample} for example images. In addition to the images, the following metadata was collected and published: 
\begin{itemize}
    \item timestamp when the image was taken,
    \item level of physical activity 1 hour prior to taking the image,
    \item temperature at the time of taking the image,
    \item whether lotion or make-up was applied at the time of taking the image, and
    \item whether the treatment was applied or not.
\end{itemize}

\section{Methods}
\subsection{Overview}

In this section, we describe different analyses of the acne image N-of-1 trial data described in the previous section. To determine whether the acne treatment was effective, we need a quantifiable way to assess each image in order to compare the images taken during treatment to those taken during no treatment. For data preparation, we preprocessed the images by resizing them to standard $224\times 224$ pixels. Then, we first performed a manual expert analysis by rating the images in their acne severity followed by standard statistical hypothesis tests on these ratings. Second, we performed an automated analysis consisting of pretraining a convolutional network on external data, applying the network on the collected images to obtain scores, and then performing standard statistical models on these scores. We hypothesized that the automated CNN approach would achieve similar results compared to the manual human labeller approach. In all statistical hypothesis tests, we considered an $\alpha$ level of 0.05.


\subsection{Manual Expert Analysis}\label{expertmodel}

\begin{figure*}[h]
    \centering
        \includegraphics[scale=0.3]{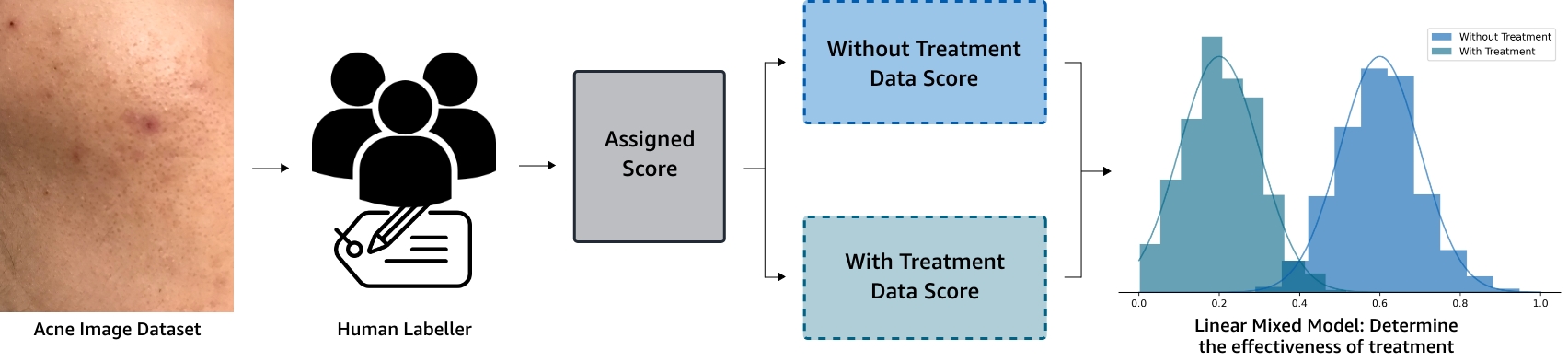}
      \caption{Illustration of the manual expert analysis.}\label{baselinefig}
\end{figure*}

An overview of the manual expert analysis is given in Figure \ref{baselinefig}. In a first step, we labeled all images regarding their acne severity. Each image was rated on a continuous scale of 0 to 1 by five raters (JF, SL, SD, SR, XG), where a value of 0 would mean no visible acne and a value of 1 would mean the skin is fully covered with acne. This process was completely blinded, with the sequence of images randomly shuffled and all metadata hidden away. No rater was able to see the ratings from any other rater. We then averaged the resulting scores into one score for each image and matched it back to the collected data. All subsequent data analyses were based on these generated labels.
Next, we performed min-max scaling of the scores given by each rater. The motivation for this approach was that the scale of scoring might be different for each rater so that it might be important to standardize across all the scores. At this point, each image is given 5 scaled scores, one from each rater. 

To evaluate whether there was a treatment effect of the acne creams with respect to the scaled scores, we fitted a linear mixed model with autoregressive errors with AR1 structure, separately for each study participant, to account for the dependence of images within each participant. 
In the linear mixed model, 
we used the average of all 5 scores of an image as response variable and the binary treatment/no-treatment variable as fixed effect. After fitting the model for each participant, we then extracted the $\beta$ coefficient as estimate of the treatment effect and performed standard Wald tests using the lme function in the nlme package in R with default settings.


\subsection{Automated CNN Analysis}

\begin{figure*}[h]
    \centering
        \includegraphics[scale=0.3]{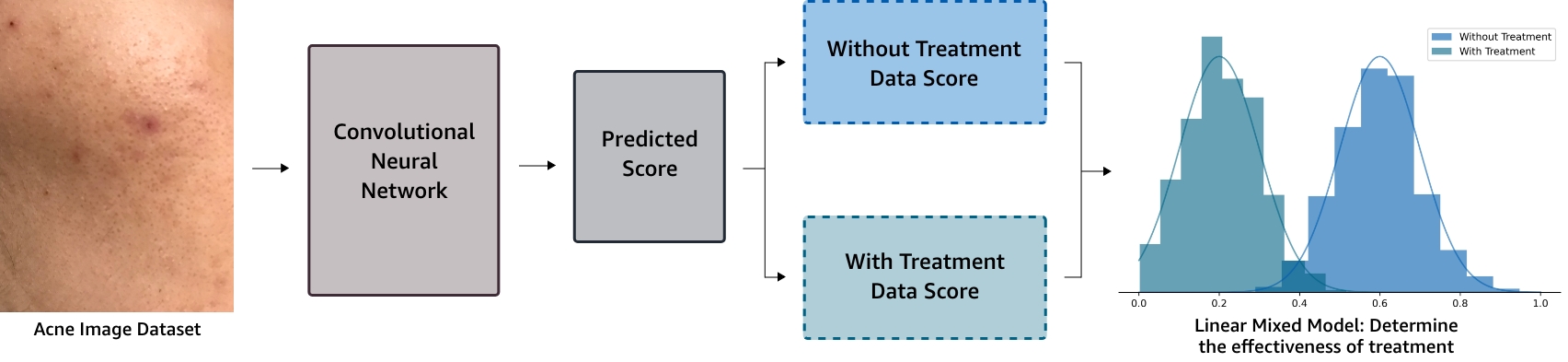}
      \caption{Diagram of CNN model pipeline.}\label{cnnfig1}
\end{figure*}

\begin{figure*}[h]
    \centering
        \includegraphics[scale=0.21]{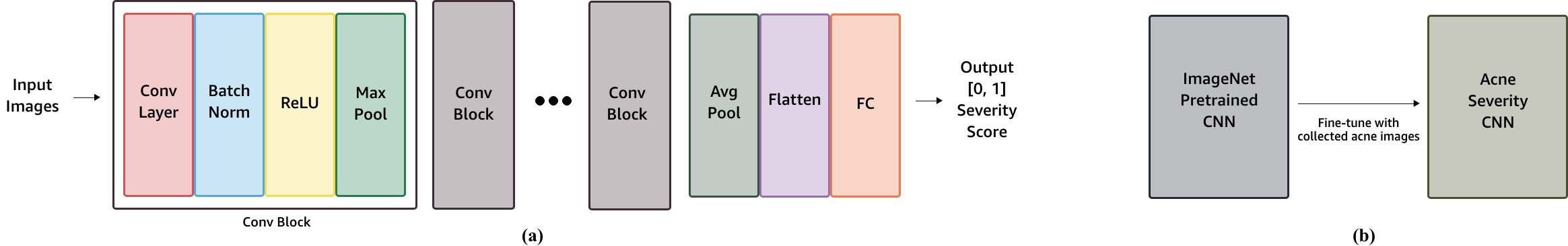}
      \caption{Illustration of (a) the CNN general structure and (b) fine-tuning.}\label{cnnfig2}
\end{figure*}

In the following, we present our proposed framework that allows to automate the scoring process and to run automated analyses and inference on image N-of-1 trials on digital platforms such as StudyU. In this approach, we extract essential features from the images by leveraging machine learning techniques instead of manual labeling, in particular, we use CNNs for our task. See Figure \ref{cnnfig1} and Figure \ref{cnnfig2} for an overview.

\subsubsection{Label Preprocessing}
For the CNNs, we performed additional preprocessing of the scores given to images. 
We gave each image a more definitive evaluation of either '0' for 'absence of acne' or '1' for 'presence of acne'. We achieved this by first taking the median of each rater’s scores and denoting all scores below the median to be 0 and scores above the median to be 1. Now, with each image having 5 binary scores, we assigned a single score to each image by taking the mode. With those new scores as targets, the problem had been reduced to a binary classification problem that we approached using a CNN. All the Python scripts for data preprocessing and the modeling are available at \url{https://github.com/HIAlab/Multimodal_Nof1/tree/main/Imaging_Nof1_trial/code}.

\subsubsection{Baseline CNN Model}
\label{sec:baseline_cnn}
We used Resnet50 \cite{he2016deep} as a pre-trained network and concatenated a fully-connected layer at the end for obtaining probability scores of having acne. The weights of the pretrained convolutional neural network were obtained by training on the benchmark ImageNet dataset. Specifically, we used the IMAGENET1K\_V2 weights in the torchvision package as initial weights of the model. We then concatenated the fully-connected layer with the available metadata such as temperature, workout level, etc. We applied a sigmoid function to the output of the fully-connected layer to obtain probability scores from 0 to 1. We used an Adam optimizer with 0.001 as our learning rate and binary cross entropy loss as our learning objective. In terms of train, validation and test split, we used one subject's images entirely for test and divided the rest of the images randomly for training and validation by 80/20 with shuffling. We used a batch size of 32, and trained the network until validation error no longer decreased. We made predictions on each image in the test set, and each score corresponds to the probability of having acne. Finally, for testing the treatment effect, we fitted the same linear mixed model as described in section \ref{expertmodel} on the probability scores extracted from the CNN.

\subsubsection{CNN with Data Augmentation}
\label{sec:cnn_data_augmentation}

Even with all images of all participants combined, only 255 images were available. In order to accommodate this limited training data, we expanded the existing dataset by data augmentation techniques and a more carefully specified train-test split. 

In more detail, similarly to the baseline CNN model described in the previous section, we included the images of 4 respective participants in the training and validation data with a random 80/20 split, and the images of the respective participant of interest in the test set. However, now, we augmented all of the training data by random rotations, horizontal flips, and brightness changes. In addition, we included augmented images of the test participant in the training and validation data. The resulting pipeline enabled us to train our model to be more robust with respect to randomness in lighting, angle, composition of the image etc. Our motivation for this configuration was the following: in a real-world scenario where you typically have access to a large amount of data for training instead of the 255 images at our current disposal, the test subject's images will inevitably resemble variations of the features in our training images. And our configuration will simulate this effect in a quantitative manner. 

After training and validation, we applied the network similarly to described in section \ref{sec:baseline_cnn}.
Finally, for testing the treatment effect, we fitted the same linear mixed model as described in section \ref{expertmodel} on the probability scores extracted from the CNN.

\subsubsection{Sensitivity Analyses}
In order to evaluate the validity of our suggested approach to incorporate augmented images of the test participant into the training, we performed two sensitivity analyses. For these analyses, we focused on participant 2.

In a first sensitivity analysis, we only included augmented images of half of the test participant's images into the training, and tested on the other half of the participant's images. This setup eliminated any overlap between training and test but effectively decreased the size of our training set. We sufficiently trained our network with this smaller dataset and predicted on the other half of the images. We then fitted a linear mixed model on the predictions and compared the p-values with those produced by the approach proposed in section \ref{sec:cnn_data_augmentation}. 

For a second sensitivity analysis, we performed a permutation simulation study. We ran 1000 inferences with our original training setup described in in section \ref{sec:cnn_data_augmentation}, i.e. with the same network architecture, training split and level of augmentation. In each iteration, we randomly permuted the assignment of images to treatment/no treatment of the test subject and fitted a linear mixed model on the predicted scores. Finally, we obtained an empirical estimate of the type I error by dividing the total number of p-values smaller than 0.05 by 1000, the number of iterations. If there was any inflation of the type I error, we would expect p-values smaller than 0.05 in a high number of iterations.

\section{Results}


\subsection{Manual Expert Analysis Identifies Treatment Effects}

First, we investigated the the results of the manual expert analysis, which are presented in Table \ref{tbl1}. The results indicated a treatment effect for two individuals, participant 1 and 4. Both participants compared treatment B to no treatment and both participants showed a positive treatment effect, i.e. decrease in acne severity scores. Interestingly, the treatment effect estimate for participant 1, who also applied treatment B, was in opposite direction but the hypothesis test did not provide evidence for a treatment effect. There was no indication of a treatment effect for participants 3 and 5 who applied treatment A.  

\begin{table}[pos=h]
\caption{Results of the manual expert analysis: estimates of the regression coefficients and p-values of their hypothesis tests based on the linear mixed model with autoregressive errors and AR1 correlation structure.}\label{tbl1}
\begin{tabular*}{\tblwidth}{@{}LCL@{}}
\toprule
& $\hat{\beta}$ & p-value \\ 
\midrule
Subject 1 & {~}0.122 & 0.10 \\
Subject 2 & -0.179 & 0.0009 \\
Subject 3 & -0.034 & 0.55 \\
Subject 4 & -0.152 & 0.03 \\
Subject 5 & -0.017 & 0.72 \\
\bottomrule
\end{tabular*}
\end{table}

\subsection{Baseline CNN Model does not Identify Treatment Effects}
The results of the baseline CNN pipeline without data augmentation are presented in Table \ref{tbl2}. While all treatment effect estimates were in the same direction as the manual expert analysis, the effect estimates were much smaller and none of the p-values of the hypothesis tests was smaller than 0.05. This suggests that the baseline CNN pipeline failed to capture the treatment effect and did not yield efficient estimates.

\begin{table}[pos=h]
\caption{Results of for the baseline CNN model: estimates of the regression coefficients and p-values of their hypothesis tests based on the linear mixed model with autoregressive errors and AR1 correlation structure on the probability scores output by the baseline CNN model.}\label{tbl2}
\begin{tabular*}{\tblwidth}{@{}LCL@{}}
\toprule
& $\hat{\beta}$ & p-value \\ 
\midrule
Subject 1 & {~}0.056 & 0.34 \\
Subject 2 & -0.123 & 0.09 \\
Subject 3 & -0.092 & 0.17 \\
Subject 4 & -0.013 & 0.88 \\
Subject 5 & -0.008 & 0.93 \\
\bottomrule
\end{tabular*}
\end{table}

\subsection{CNN with Data Augmentation Identifies Treatment Effects}
Our proposed framework described in section \ref{sec:cnn_data_augmentation} used data augmentation in order to include more data into training and produce a more robust model. The results of this framework are shown in Table \ref{tbl3}. Again, all treatment effect estimates were in similar direction compared to the manual expert analysis (except for participant 5, who had treatment effect estimates very close to 0). However, in this analysis, the treatment effect for participant 2 was estimated to be even larger, and the p-value was similar compared to the manual expert analysis. There was no indication of a treatment effect for participant 4.

\begin{table}[pos=h]
\caption{Results of for the CNN model with data augmentation: estimates of the regression coefficients and p-values of their hypothesis tests based on the linear mixed model with autoregressive errors and AR1 correlation structure on the probability scores output by the CNN model with data augmentation.}\label{tbl3}
\begin{tabular*}{\tblwidth}{@{}LCL@{}}
\toprule
& $\hat{\beta}$ & p-value \\ 
\midrule
Subject 1 & {~}0.050 & 0.61 \\
Subject 2 & -0.278 & 0.002 \\
Subject 3 & -0.017 & 0.84 \\
Subject 4 & -0.015 & 0.87 \\
Subject 5 & {~}0.026 & 0.76 \\
\bottomrule
\end{tabular*}
\end{table}

\subsubsection{Sensitivity Analyses Support Validity of Framework}
In the first sensitivity analysis, we trained the augmented CNN network with half of the images of participant 2 and left the other half as test set. We trained the network sufficiently until validation error no longer decreased. The subsequent test based on the linear mixed model, whether the predicted probabilities of acne severity differ between treatment and no treatment, yielded a p-value of 0.07.
This p-value was slightly larger compared to what we obtained from our original augmented setup which, in our opinion, supports the validity of our approach in that there still seemed to be some signal but without enough power.

In the second sensitivity analyses, we performed a simulation study to estimate the empirical type I error of our proposed augmentation approach. For this, we permuted the treatment and no-treatment sequence for the test subject for every inference hence performed hypotheses tests under the null hypothesis. Across the 1000 iterations, 49 out of 1000 p-values were smaller than 0.05. That is, our estimate for the empirical type I error was 0.049 which was almost identical to the desired 0.05 level.

Both experiments provide support for the validity of our proposed automated augmented CNN framework. 

\section{Discussion}
In this study, we have proposed a novel framework for multimodal N-of-1 trials. Multimodal N-of-1 trials can be the tool for fully using all capabilities of modern smartphone technology by incorporating health outcomes assessed through video, audio or pictures. We illustrate the framework for a series of N-of-1 trials on acne treatments where the outcome was captured by images. In the analysis of the data, we perform a traditional manual analysis as baseline that would be typically applied in practise by a physician or researcher. Importantly, we then propose and demonstrate how such data can also be automatically analyzed using deep learning architectures. Such automated data analysis approaches can allow to widely apply multimodal N-of-1 trials and our results showed that they are able to replicate the effects observed in the expert analysis.
Very interestingly, the results of the analysis were able to identify a positive treatment effect in one participant, despite the short length of the N-of-1 trial with 16 days. It would be interesting to follow-up this observation in a larger acne image trial that investigates the treatment effects over a longer time frame and with longer treatment blocks.

There are several limitations that we want to acknowledge in our study. First, even with a standardized protocol across the five participants, the image quality could still vary, which could not only impact our human labeled scores but also the modeling accuracy. The identification of a treatment effect in one participant suggests that there was sufficient data quality. Further standardization might be important, of the aim of a series of multimodal N-of-1 trials lies as well on an aggregated analysis of the data. Second, for the current model to generalize well, it requires a sufficient amount of labeled data for training. We were able to generate sufficient training data in our proposed augmented approach and by clever inclusion of augmented images into the training. Future research can investigate unsupervised approaches in order to reduce the necessary labeling for the analysis.

Our analysis has a wide range of future directions. First, in addition to investigating alternative modeling approaches of the images mentioned above, our proposed pipeline might be improved by including transfer learning approaches or including bounding boxes to the treatment area in order to help standardising the image dimensions. Second, future work can focus on integrating multimodal N-of-1 trials and their automated analyses into the existing StudyU platform. Third, incorporating more modalities such as audio and video data, can further extend the range of possibilities and make N-of-1 trials more accessible. 

Overall, our work contributes to the ongoing efforts to make healthcare more accessible and customized to individual needs. Multimodal N-of-1 trials have the potential to greatly improve the accessibility and effectiveness of personalized healthcare. By gathering more detailed and comprehensive data about individual patients and using this information to create personalized treatment plans, these trials can help to ensure that patients receive the most appropriate care for their unique needs and circumstances.

\printcredits

\bibliographystyle{cas-model2-names}

\bibliography{Paper}

\bio{}
\endbio

\newpage

\beginsupplement

\section*{Supplementary Material}

\subsection*{Supplementary Tables}

\begin{table}[pos=h]
    \centering
    \begin{tabular}{|c|ccccc|}
         \hline
         Participant & 1 & 2 & 3 & 4 & 5 \\
         \hline
         Day 1 & / & / & / & / & / \\
         Day 2 & / & / & / & / & / \\
         Day 3 & B & B & A & B & A \\
         Day 4 & B & B & A & B & A \\
         Day 5 & / & / & / & / & / \\
         Day 6 & / & / & / & / & / \\
         Day 7 & B & B & A & B & A \\
         Day 8 & B & B & A & B & A \\
         Day 9 & / & / & / & / & / \\
         Day 10 & / & / & / & / & / \\
         Day 11 & B & B & A & B & A \\
         Day 12 & B & B & A & B & A \\
         Day 13 & / & / & / & / & / \\
         Day 14 & / & / & / & / & / \\
         Day 15 & B & B & A & B & A \\
         Day 16 & B & B & A & B & A \\
         \hline
    \end{tabular}
    \caption{Complete Schedule of Treatment: A and B stand for receiving treatment of type A and B. Days without treatment are symbolized with \texttt{/}.}
    \label{tab:schedule}
\end{table}

\newpage

\subsection*{Supplementary Figures}
\begin{figure}[h]
    \centering
    \subfloat[\centering Day 1 Sample]{{\includegraphics[width=0.22\textwidth]{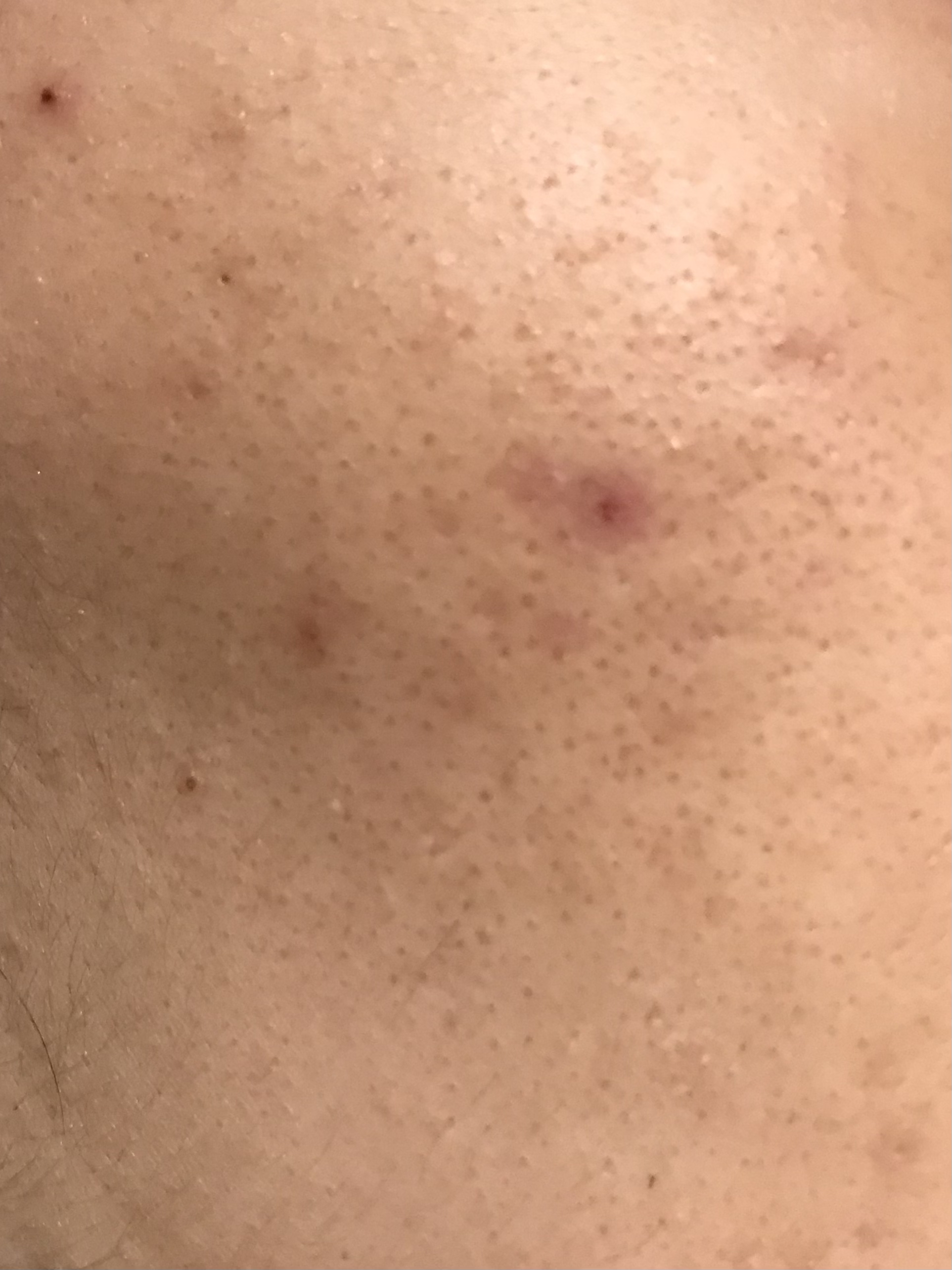} }}%
    \subfloat[\centering Day 16 Sample]{{\includegraphics[width=0.22\textwidth]{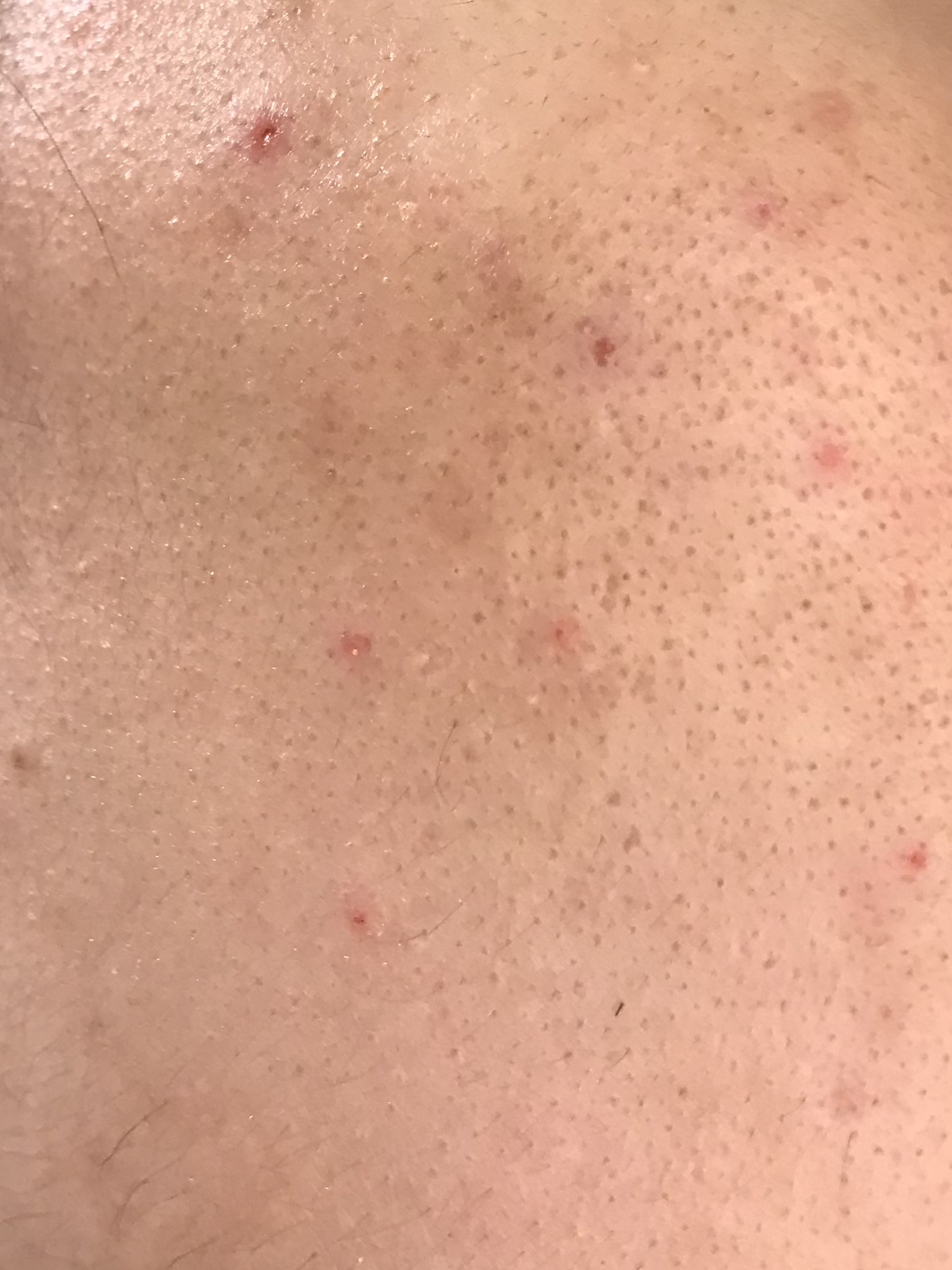} }}%
    \caption{Image Samples of Participant 1.}%
    \label{fig:xg-sample}
\end{figure}

\begin{figure}[h]
    \centering
    \subfloat[\centering Day 1 Sample]{{\includegraphics[width=0.22\textwidth]{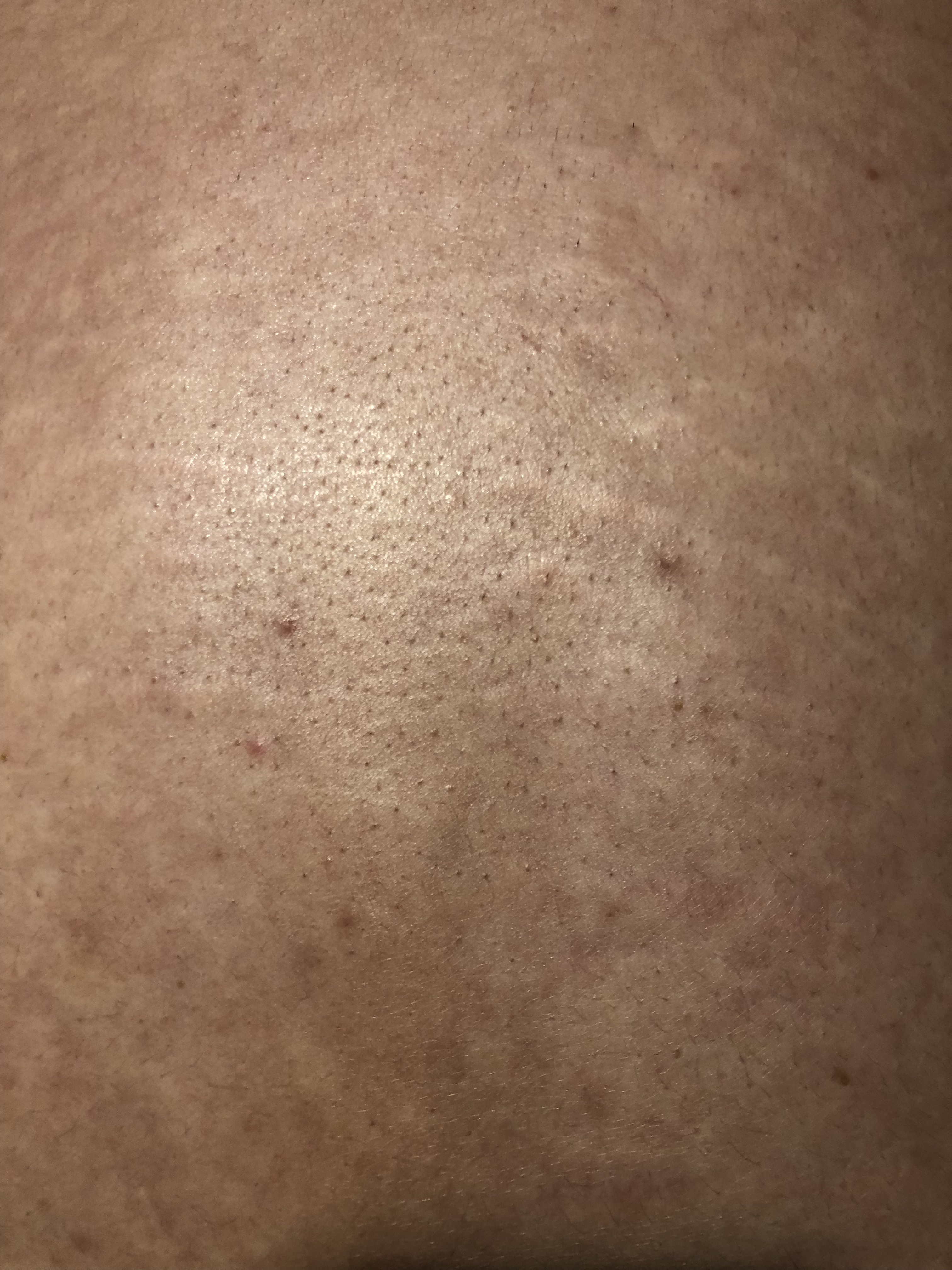} }}%
    \subfloat[\centering Day 16 Sample]{{\includegraphics[width=0.22\textwidth]{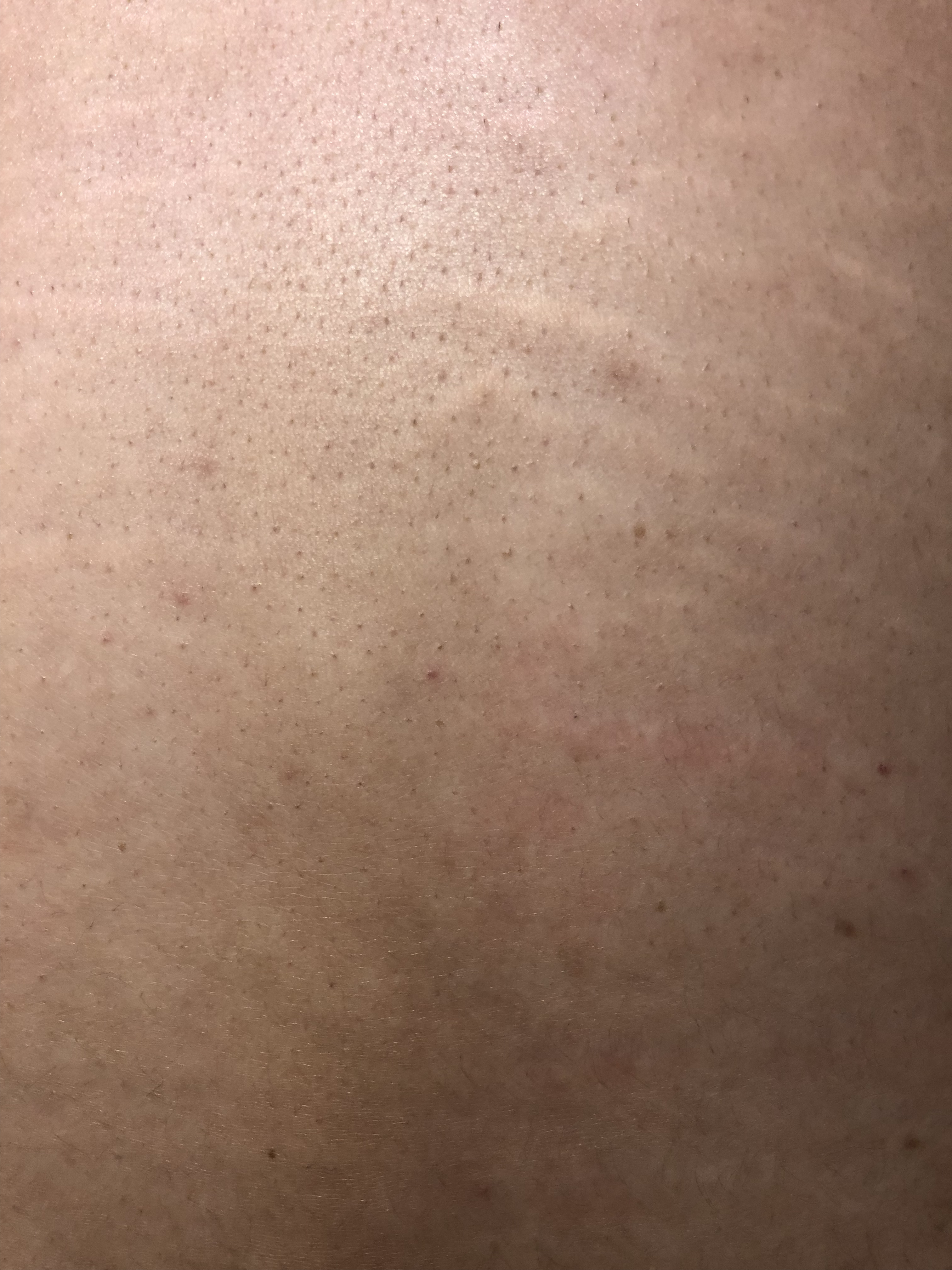} }}%
    \caption{Image Samples of Participant 2.}%
    \label{fig:sr-sample}
\end{figure}

\begin{figure}[h]
    \centering
    \subfloat[\centering Day 1 Sample]{{\includegraphics[width=0.22\textwidth]{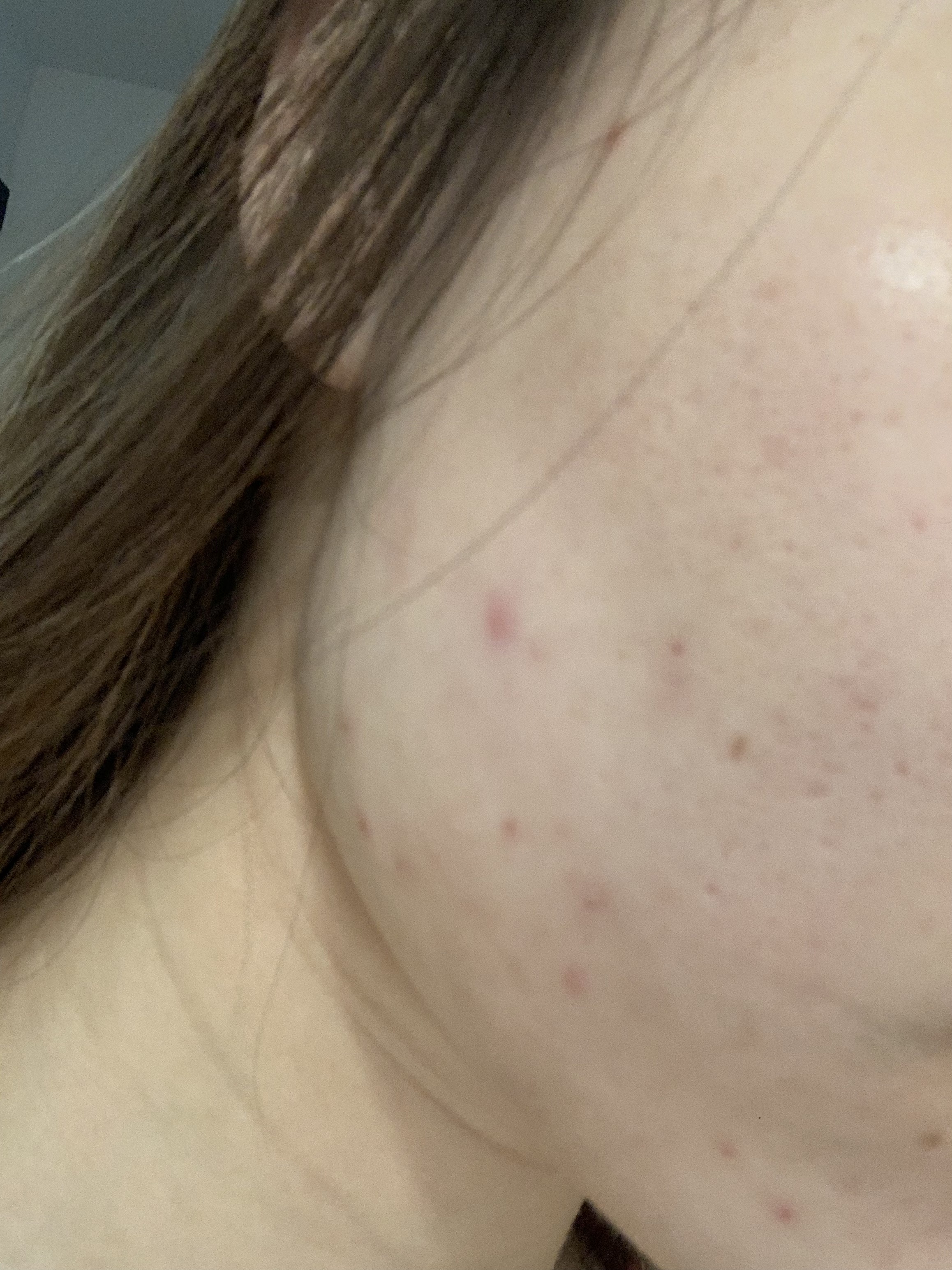} }}%
    \subfloat[\centering Day 16 Sample]{{\includegraphics[width=0.22\textwidth]{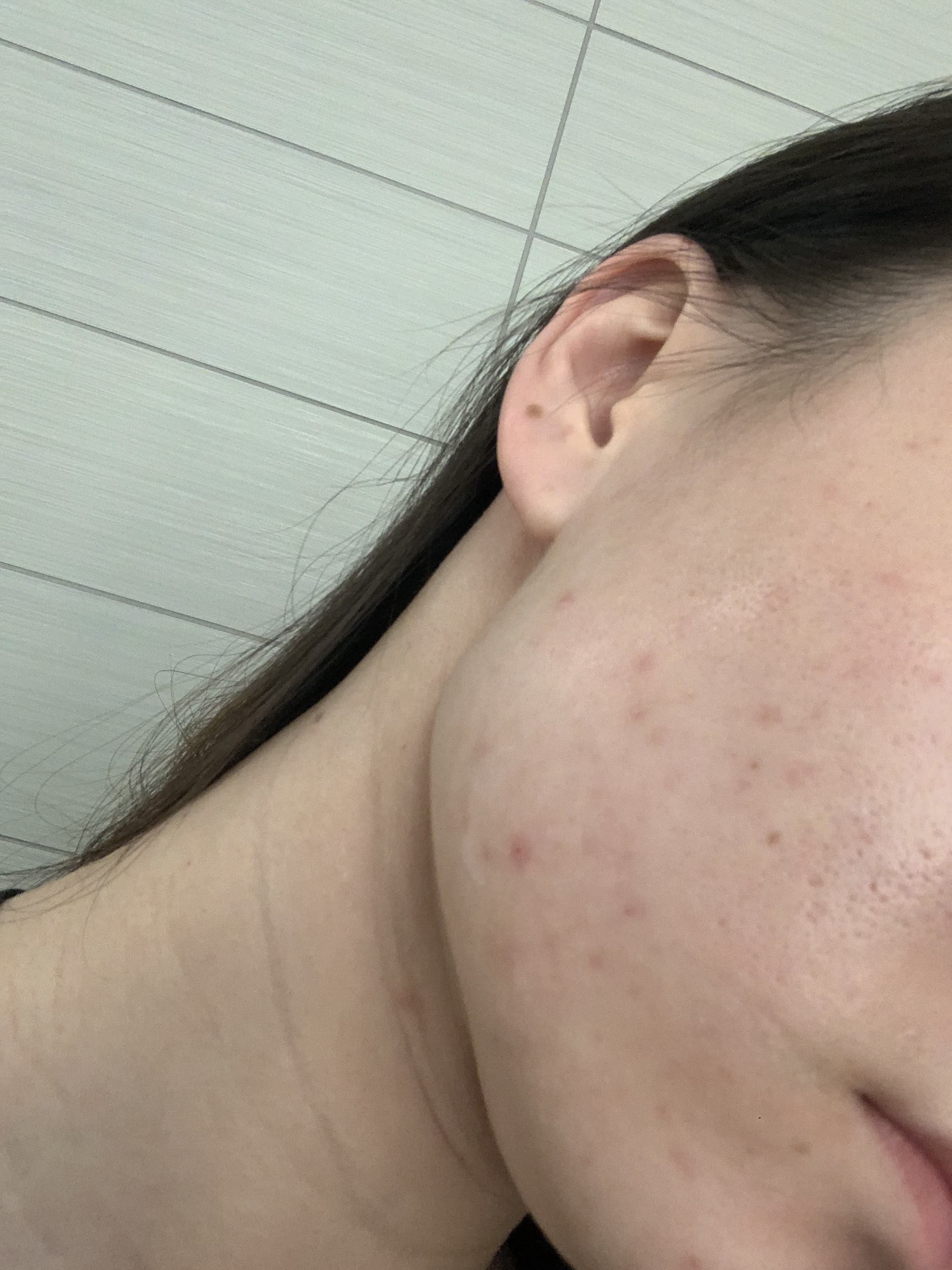} }}%
    \caption{Image Samples of Participant 3.}%
    \label{fig:jf-sample}
\end{figure}

\begin{figure}[h]
    \centering
    \subfloat[\centering Day 1 Sample]{{\includegraphics[width=0.22\textwidth]{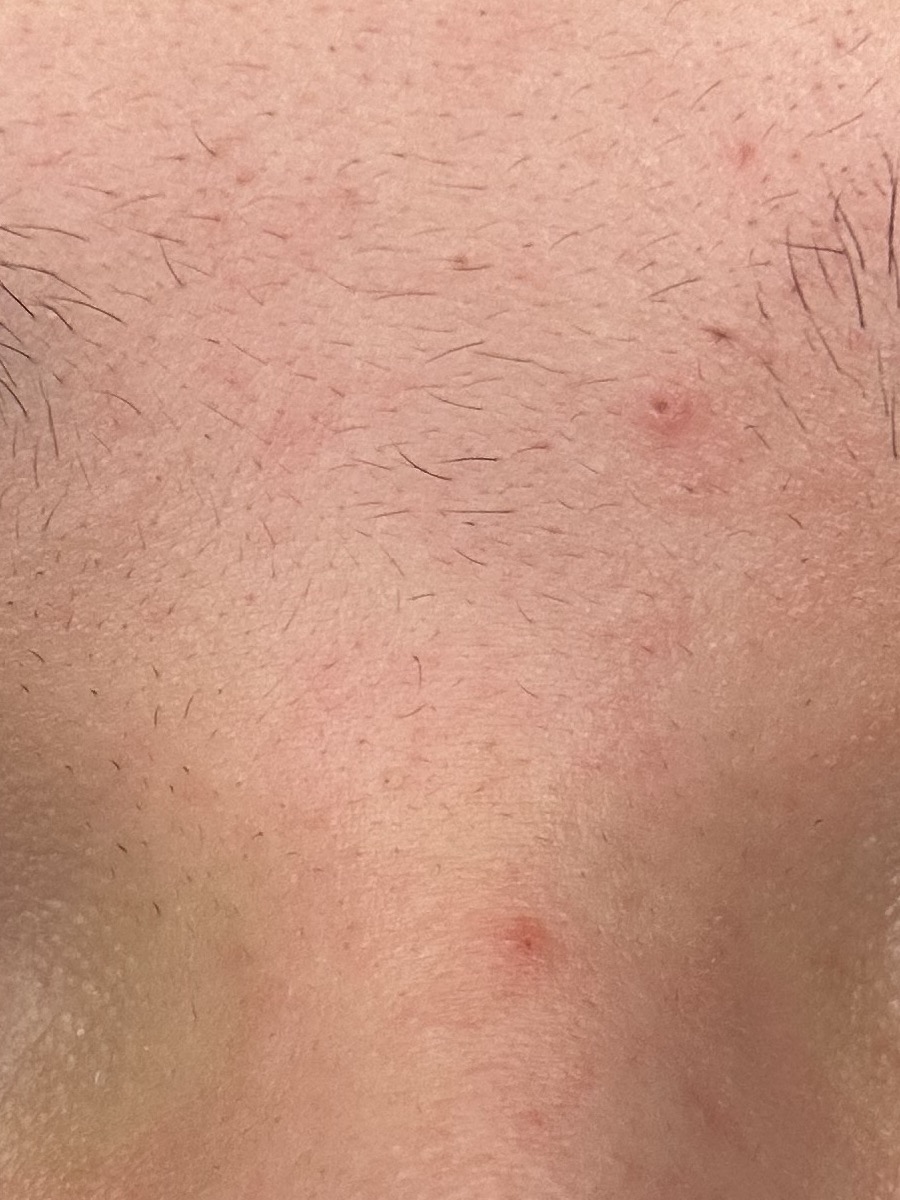} }}%
    \subfloat[\centering Day 16 Sample]{{\includegraphics[width=0.22\textwidth]{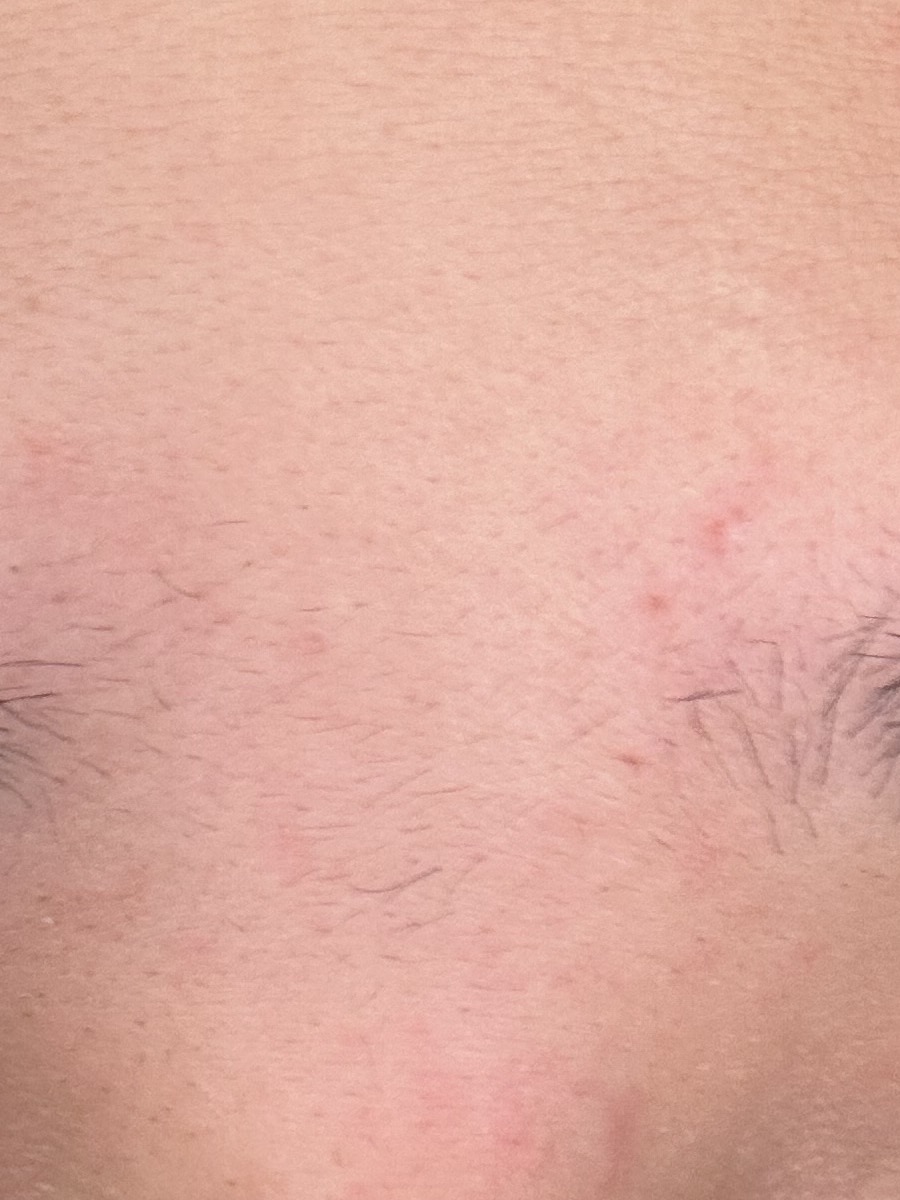} }}%
    \caption{Image Samples of Participant 4.}%
    \label{fig:sl-sample}
\end{figure}

\begin{figure}[h]
    \centering
    \subfloat[\centering Day 1 Sample]{{\includegraphics[width=0.22\textwidth]{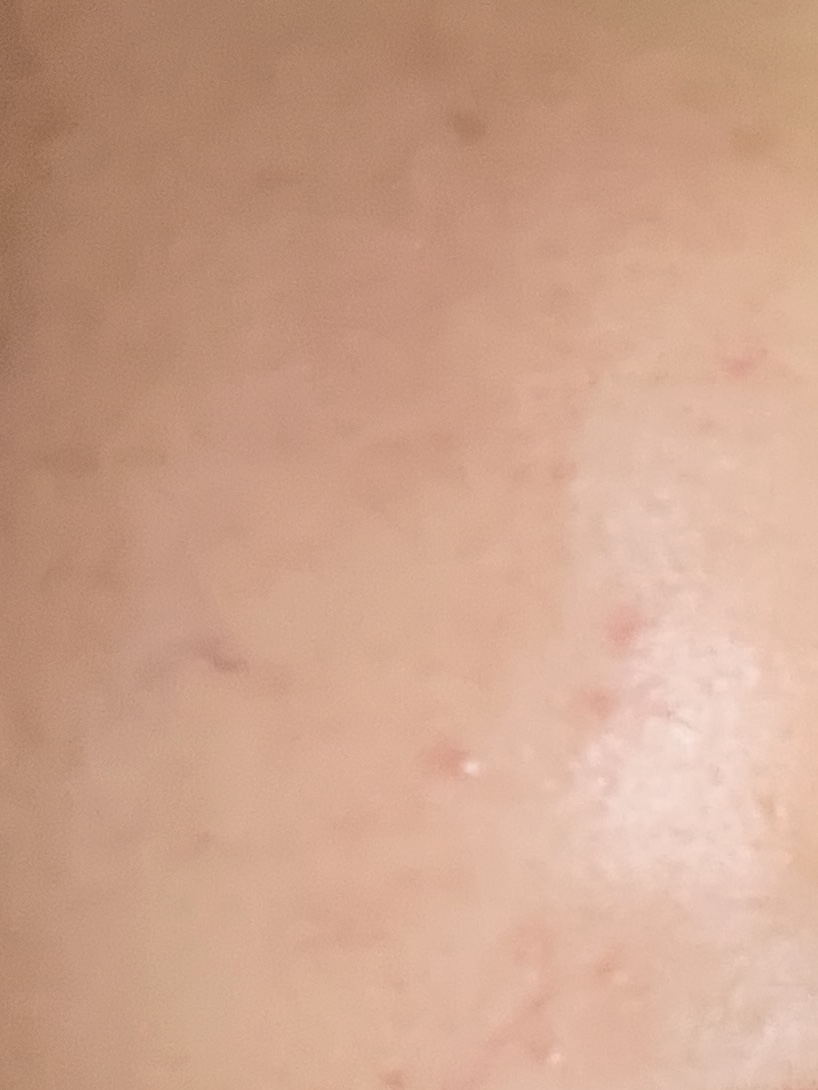} }}%
    \subfloat[\centering Day 16 Sample]{{\includegraphics[width=0.22\textwidth]{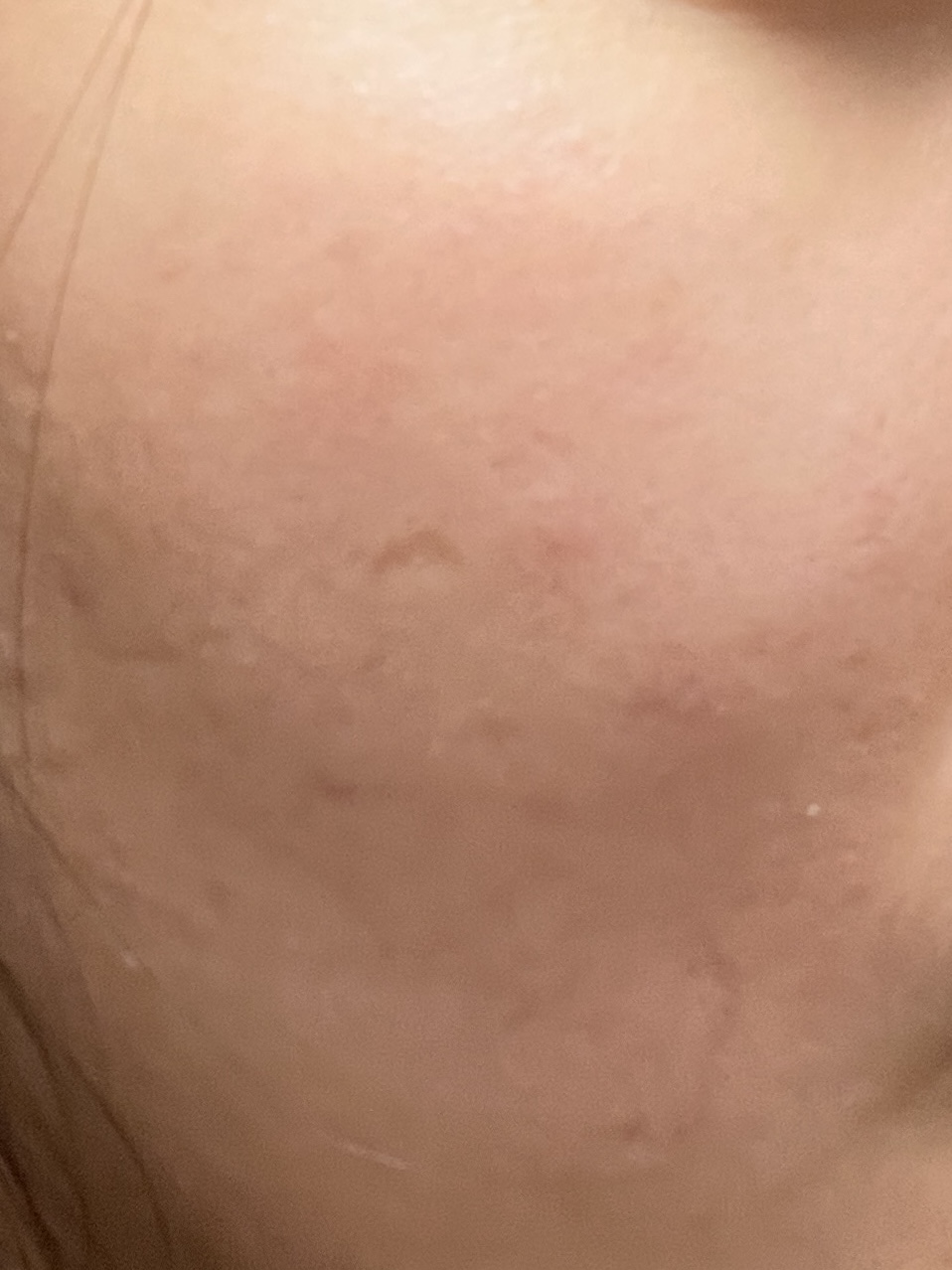} }}%
    \caption{Image Samples of Participant 5.}%
    \label{fig:sd-sample}
\end{figure}

Two image samples of each study participant can be found in Figures \ref{fig:xg-sample} - \ref{fig:sd-sample}.


\end{document}